\documentclass[twocolumn,letter]{IEEEtran}
\usepackage[usenames,dvipsnames]{color}
\usepackage{subfigure}
\usepackage{graphicx}
\usepackage{amsmath}
\usepackage{color}
\usepackage{multicol}
\usepackage{amssymb}

\usepackage{amsfonts}%
\usepackage{verbatim}%
\usepackage{enumerate}%
\usepackage{psfrag}
\usepackage{epsfig}
\usepackage{multicol}
\usepackage{setspace}
\usepackage{dsfont}
\usepackage{balance}

\usepackage{algorithm}
\usepackage{algorithmic}
\usepackage{cite}
\usepackage{float}
\usepackage{cite}
        
 \newtheorem{proposition}{Proposition}
\begin{document}
\title{Maximum Throughput of a Cooperative Energy Harvesting Cognitive Radio User}
\author{Ahmed~El~Shafie,~\IEEEmembership{Member,~IEEE,}
        Tamer~Khattab,~\IEEEmembership{Member,~IEEE,}
        ~and H. Vincent Poor,~\IEEEmembership{Fellow,~IEEE}
        \thanks{Part of this paper has been accepted in Personal, Indoor and Mobile Radio Communications (PIMRC), 2014 \cite{myprotocol}.}
\thanks{A. El Shafie is with Wireless Intelligent Networks Center (WINC), Nile University, Giza, Egypt. He is also with Electrical Engineering, Qatar University, Doha, Qatar (e-mail: ahmed.salahelshafie@gmail.com).}
\thanks{T. Khattab is with Electrical Engineering, Qatar University, Doha, Qatar (email: tkhatta@ieee.org).}
\thanks{H. V. Poor
is with the Department of Electrical Engineering, Princeton University, Princeton, NJ 08544 USA (email:
poor@princeton.edu).}
\thanks{This research work is supported by Qatar National Research Fund (QNRF) under grant number NPRP 09-1168-2-455.}}
\date{}
\maketitle
\begin{abstract}
In this paper, we investigate the maximum throughput of a saturated rechargeable secondary user (SU) sharing the spectrum with a primary user (PU).
The SU harvests energy packets (tokens) from the environment with a certain harvesting rate. All transmitters are assumed to have data buffers to store the incoming data packets. In addition to its own traffic buffer, the SU has a buffer for storing the admitted primary packets for relaying; and a buffer for storing the energy tokens harvested from the environment. We propose a new cooperative cognitive relaying protocol that allows the SU to relay a fraction of the undelivered primary packets. We consider an interference channel model (or a multipacket reception (MPR) channel model), where concurrent transmissions can survive from interference with certain probability characterized by the complement of channel outages. The proposed protocol exploits the primary queue burstiness and receivers' MPR capability. In addition, it efficiently expends the secondary energy tokens under the objective of secondary throughput maximization. Our numerical results show the benefits of cooperation, receivers' MPR capability, and secondary energy queue arrival rate on the system performance from a network layer standpoint.
\end{abstract}
\begin{IEEEkeywords}
Cognitive radio, relaying, protocol, cooperation, throughput analysis, queue stability.
\end{IEEEkeywords}

\section{Introduction}
  \IEEEPARstart{S}\small{econdary} utilization of a licensed primary band can efficiently enhance the spectrum usage and improve its scarcity. Secondary users (SUs) can use the spectrum under certain quality of service requirements for the primary users (PUs). High performance wireless communication networks relies, among other technologies, on cooperative communications, where nodes cooperate to mitigate fading.

In many practical situations and applications in wireless sensor networks, the SU is a battery operated device. The secondary operation, which involves spectrum sensing and access, is accompanied by energy consumption. Consequently, an energy-aware (energy-efficient) SU must optimize its sensing and access decisions to efficiently invest the available energy. When the SU is capable of relaying, it should also optimize its decision on accepting other nodes packets for relaying. This is because accepting a packet for relaying will require its retransmission and therefore consumes energy.

Energy harvesting technology is an emerging technology for energy-constrained terminals which allows the transmitter to collect (harvest) energy from its environment. For a comprehensive overview of the different energy harvesting technologies, the reader is referred to \cite{survey} and the references therein.

Data transmission by an energy harvester with a rechargeable battery has got a lot of attention recently  \cite{lei2009generic,sharma2010optimal,gatzianas2010control,pappas2012optimal,wimob,ourletter,ElSh1312:Optimal,krikidis2012stability,wcmpaper}.
 In \cite{lei2009generic}, the optimal
online policy for controlling admissions into the data buffer
is derived using a dynamic programming framework. In \cite{sharma2010optimal},
energy management policies which stabilize the data queue
are proposed for single-user communication and some delay-optimal properties are derived.
In \cite{gatzianas2010control}, the optimality of a variant of the back-pressure algorithm
using energy queues is shown.

The authors of \cite{pappas2012optimal} considered a cognitive scenario where two different priority nodes share a common channel. The higher priority user (PU) has a rechargeable battery, whereas the lower priority user (SU) is plugged to a reliable power supply and therefore has energy each time slot without limitations. In \cite{wimob}, the authors investigated a cognitive setting with one PU and one rechargeable SU. The SU randomly accesses and senses the primary channel and can possibly leverage primary feedback. Receivers are capable of decoding under interference as they have multipacket reception (MPR) capabilities. The authors investigated the maximum secondary throughput under stability and delay constraints on the primary queue. In \cite{ourletter}, the SU randomly accesses the channel at the beginning of the time slot to exploit the MPR capability of receivers. The SU aims at maximizing its throughput under stability and queueing delay constraints on the primary queue. In \cite{ElSh1312:Optimal}, El~Shafie {\it et al.} investigated the maximum stable throughput of an energy harvesting SU under stability of an energy harvesting primary transmitter. The SU selects a sensing duration each time slot from a predefined set such that its stable throughput is maximized under the stability of the primary queue.

Cooperative cognitive relaying has got extensive attention recently \cite{sadek,simeone,krikidis2009protocol,krikidis2012stability,wcmpaper}. 
In \cite{sadek}, Sadek {\it et al.} proposed cognitive protocols for a multiple access system with a single relay that aids the transmitting nodes. The proposed cooperative protocols enable the relaying node to help a set of buffered transmitters operating in a time-division multiple access network when their queues are empty due to
source burstiness. The secondary throughput of the proposed protocol as well as the delay of symmetric nodes were investigated. The authors of \cite{simeone} investigated a network composing of one primary transmitter-receiver pair and one secondary transmitter-receiver pair. The cognitive radio transmitter aims at maximizing its throughput via optimizing its transmit power such that the primary and the relaying queues are maintained stable.

Integrating cooperative communications and energy harvesting technologies has been considered in several works such as \cite{krikidis2012stability} and \cite{wcmpaper}. In \cite{krikidis2012stability}, the authors investigate the effects of network layer
cooperation in a wireless three-node network with energy harvesting
nodes and bursty data traffic. The authors derived the maximum
stable throughput of the source as well as
the required transmitted power for both a non-cooperative and an
orthogonal decode-and-forward cooperative schemes. In \cite{wcmpaper}, the authors study the impact of the energy queue on the maximum stable throughput of a cooperative energy harvesting SU that utilizes the spectrum whenever the PU's queue is empty. The authors assume an energy packet consumption in either data decoding or data transmission. Inner and outer bounds are derived for the secondary throughput

In this work, we investigate the maximum throughput for an energy harvesting SU in presence of a PU. In contrast to \cite{wcmpaper} and \cite{krikidis2012stability}, we consider a generalized MPR channel model and propose a new cooperative protocol which exploits the MPR capability of the receivers. In the proposed cooperative cognitive relaying protocol, the SU cooperatively relays a fraction of undelivered primary packets. The flow of the primary packets through the SU's relaying queue is controlled using some tunable parameters which depend on the channels quality and other queues states. The proposed cooperative cognitive relaying protocol allows the SU to transmit simultaneously with the PU at a fraction of the time slots to exploit the MPR capability of the receiving nodes. The proposed protocol is simple and doesn't require continuous estimation of the channel state information (CSI) at the transmitting terminals.

The contributions of this paper can be summarized as follows:
\begin{itemize}
\item We propose a new cooperative cognitive relaying protocol which exploits primary queue burstiness and receivers MPR capability while efficiently managing the SU's energy expenditure.
    \item We derive the channel outages probabilities in case of energy harvesting nodes and the presence of delays in transmission with and without concurrent transmissions. Moreover, we shed light on some important issues related to access delays on channel outages with and without concurrent transmissions.
     \item We derive service and arrival mean service rates expressions.
    \item We investigate the maximum throughput of the SU under the stability of all other queues in the system.
  \item We study the impact of the MPR capability of the receivers and the secondary energy queue on the secondary throughput.
  \item To make the characterization of the secondary throughput feasible, we consider three approximated systems: two of them are shown to be inner bounds on the performance of the original system, whereas the third is shown to be an outer bound on the performance of the original system.
            \end{itemize}

            This paper is structured as follows: Next we describe the system model adopted in this paper. We explain the proposed cooperative cognitive relaying protocol and provide the analysis of the queues rates and the problem formulation in Section \ref{protocol}. In Section \ref{numerical}, we provide some numerical results. The conclusions are drawn in Section \ref{conclusion}.

\section{System Model}\label{sysmodel}
\begin{figure}
  \centering
  \includegraphics[width=1\columnwidth]{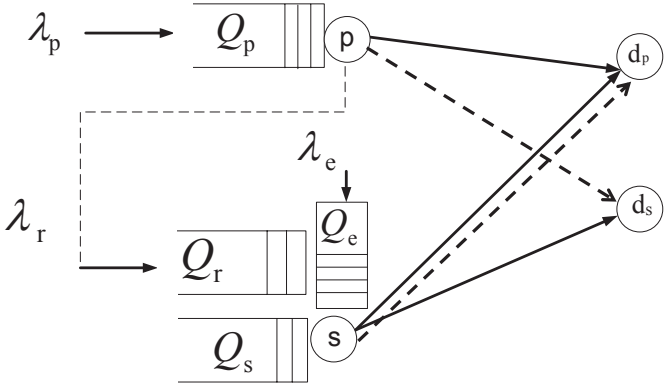}\\
  \caption{Primary and secondary queues and links. In the figure, the solid lines are the communication channels while the dashed lines are the interference channels. 
  }\label{fig1}
  \end{figure}

We consider a simple configuration comprised of one rechargeable battery operated secondary transmitter `${\rm s}$', one secondary destination `${\rm d_s}$', one primary transmitter `${\rm p}$' and one primary destination `${\rm d_p}$'. The network model is shown in Fig. \ref{fig1}. The primary transmitter-receiver pair operates over slotted channels. Time is slotted and a slot is $T$ seconds in length. Each transmitter has an infinite-length data buffer (queue) to store its own incoming fixed-length data packets, denoted by $Q_{\ell}$, $\ell\in\{\rm p,s\}$. In addition to its own traffic queue, the cognitive user has an infinite capacity buffer to store the energy packets harvested from the environment; and an infinite capacity relaying queue to store the accepted primary packets for relaying.\footnote{This is a reasonable approximation if the energy contained inside one energy packet is much less than the total capacity of the energy buffer (or the battery storage capacity) \cite{krikidis2012stability,ourletter}.}  Let $Q_{\rm r}$ denote the secondary relaying queue and $Q_{\rm e}$ denote the secondary energy queue with mean arrival rates $0\!\le\!\lambda_{\rm r}\!\le\!1$ packets/slot and $0\!\le\!\lambda_{\rm e}\!\le\!1$ energy packets/slot, respectively. The secondary data queue is assumed to be saturated (always backlogged). Arrivals at queues $Q_{\rm p}$ and $Q_{\rm e}$ are assumed to be Bernoulli random variables~\cite{krikidis2010stability,close}. The arrivals at each queue are assumed to be independent and identically
distributed (i.i.d.). The Bernoulli model is simple, but it captures the
random availability of ambient energy sources. In the analysis of discrete-time queues,
Bernoulli arrivals see time averages (BASTA) is an important feature, which is equivalent to the Poisson arrivals
see time averages (PASTA) property in continuous-time
systems \cite{ourletter}.
 Arrivals are also independent from queue to queue. The mean arrival rate to the primary queue, $Q_{\rm p}$, is $\lambda_{\rm p}\in[0,1]$ packets/slot. All data packets are of size $\mathcal{B}$ bits. We assume that one energy packet is needed for the transmission of one data packet. The energy queue has energy packets each of ${\rm e}$ energy units. For similar assumptions of infinite size of data buffers and modeling the arrivals of data and energy queues as Bernoulli arrivals, the reader is referred to \cite{pappas2012optimal,wimob,ourletter,ElSh1312:Optimal,krikidis2012stability} and the references therein.

The proposed cooperative cognitive relaying protocol and the theoretical development in this work can be readily generalized to networks with more than one PU and more than one SU, where several PUs may choose one or more SUs or the best SU for cooperation.\footnote{The considered network can be seen as part of a larger network with multiple primary nodes assigned to orthogonal frequency bands or different time slots via employing frequency division multiple-access or time division multiple-access, respectively.}


All wireless links exhibit a stationary non-selective Rayleigh block fading. The instantaneous channel fading coefficient of link $j\!\rightarrow k$ (link connecting nodes $j$ and $k$) remains constant during a time slot $\mathbb{T}\in\{1,2,3,\dots\}$, but changes independently and identically from one
slot to another according to a circularly symmetric complex
Gaussian distribution with zero mean and variance $\sigma_{jk}$. Received signals at node $k$ are corrupted by complex additive white Gaussian
noise (AWGN) with zero mean and variance $\mathcal{N}_k$ Watts.

 Let $\zeta_{jk}$ denote the fading gain of link $j\rightarrow k$. We do not assume the availability of CSI at the transmitters. Since the PU transmits from the beginning of the time slot over the whole slot duration if its queue is nonempty, the spectral efficiency of the primary terminal is $R_{\rm p}=\mathcal{B}/(TW)$ bits/sec/Hz, where $W$ is the channel bandwidth. The cognitive radio user may transmit either at the beginning of the time slot or after $\tau$ seconds from the beginning of the time slot. Hence, the secondary transmission time is $T^{\left(i\right)}_{\rm s}\!=\!T\!-\!i\tau$, where $i\!=\!0$ if the SU transmits at $t\!=\!0$, and $i\!=\!1$ if the SU transmits at $t\!=\!\tau$. The spectral efficiency of the secondary transmission is either $R^{\left(0\right)}_{\rm s}=\mathcal{B}/(TW)$ bits/sec/Hz or $R^{\left(1\right)}_{\rm s}=\mathcal{B}/((T\!-\!\tau)W)$ bits/sec/Hz for $i\!=\!0$ and $i\!=\!1$, respectively. Note that the {\it decision duration}, $\tau$, should be long enough to justify the perfect detection of the primary queue state.\footnote{We assume here that the PU and the SU dedicate a special channel with small bandwidth for sharing state information of the PU. Specifically, during the first $\tau$ seconds of the time slot, the PU cooperatively sends its own queue state, i.e., empty or nonempty, to the SU each time slot over the dedicated bandwidth. This can be done through one-bit signal sent from the PU to the SU.} The PU transmits data with a fixed power $\mathbb{P}_{\rm p}$ Watts, whereas the SU transmits with power $\mathbb{P}^{\left(i\right)}_{\rm s}\!=\!{\rm e}/T_i$ Watts, $i\in\{0,1\}$. The secondary transmit power is a function of the time instant in which the SU starts data transmission within the time slot. Outage of a link occurs
when the instantaneous capacity of that link is lower than the transmitted spectral efficiency rate \cite{krikidis2012stability,wimob,ourletter}.

Assume that node $j$ transmits a packet to node $k$ and at the same time node $v$ transmits to its respective receiver. Due to the broadcast nature of the wireless communication channel, the signal transmitted by node $v$ arrives at node $k$ and causes interference with the signal transmitted by node $j$. Let us assume that node $j$ starts transmission at $t\!=\!i\tau$, whereas node $v$ starts transmission at $t\!=\!n\tau$, where $i,n\in\{0,1\}$. Under this setting, the probability that a transmitted packet by node $j$ being successfully received at node $k$ is $\overline{P_{jk,in}^{\left({v}\right)}}\!=\!1\!-\!P_{jk,in}^{\left({v}\right)}$ (see Appendix A for the exact expression). If transmitter $j$ sends its packet alone (without interference) to node $k$, and starts transmission at $t\!=\!i\tau$, the probability of that packet being successfully decoded at $k$ is $\overline{P_{jk,i}}$. The physical layer is explained with details in Appendix~A.

 A fundamental performance measure of a communication network is the stability of its queues. Stability can be defined rigorously as follows. Denote by $Q^{\left(t\right)}$ the length of queue $Q$ at the beginning of time slot $\mathbb{T}$. Queue $Q$ with mean arrival rate $\lambda$ and mean service rate $\mu$ is said to be stable if $\lim_{\kappa \rightarrow \infty  } \lim_{\mathbb{T} \rightarrow \infty  } {\rm Pr}\{Q^{\left(\mathbb{T}\right)}<\kappa\}=1$~\cite{sadek}, where ${\rm Pr}\{.\}$ denotes the probability of the event in the argument. For strictly stationary arrival and service processes, queue $Q$ is stable if $\mu\ge \lambda$. In a multiqueue system, the system is stable when \emph{all} queues are stable.
\section{Proposed cooperative cognitive relaying protocol}\label{protocol}

\begin{figure}
  \centering
  \includegraphics[width=1\columnwidth]{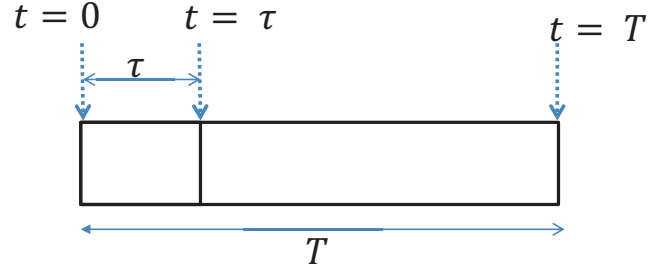}\\
  \caption{Time slot structure.
  }\label{fig0}
  \end{figure}

In this section, we describe in details the proposed cooperative cognitive relaying protocol, denoted by $\mathcal{S}$. The time slot structure is shown in Fig. \ref{fig0}.
At the beginning of the time slot, if the secondary energy queue is nonempty, the SU may decide to receive the primary packet with probability $f$ or decide to access the channel using one of its queues with probability $\overline{f}$.\footnote{Throughout this paper, $\overline{\phi}\!=\!1\!-\!\phi$.} Accessing the channel at the beginning of the time slot is motivated by the following facts:
 \begin{itemize}
 \item First, it may be the case that using the whole time in data transmission provides higher throughput than wasting $\tau$ seconds for channel sensing, specially at low primary arrival rate as the PU will be inactive during most of the time slots. Moreover, the probability of being in outage of a link decreases with the total time used in data transmission over that link. This fact is discussed and its formula is proved in Appendix~A.
      \item Second, the presence of MPR capability at the receiving nodes allows packets decoding under interference with nonzero probability, which can be exploited by the SU to boost its throughput.
          \item Third, as will be explained in details later, due to the fixed energy transmission property of the energy harvesting SU, secondary delays of channel access may increase the interference at the primary destination due to the increases of the secondary transmit power, which in turn reduces the probability of successful decoding of the primary packets at the primary destination.
              \end{itemize}
              \noindent Based on these observations, channel accessing at the beginning of the time slot may be useful for certain scenarios and under specific system and channel parameters. On the contrary, if the SU decides to receive the primary packet in a time slot, it will take another action/decision after $\tau$ seconds from the beginning of the time slot. The {\it decision duration} $\tau$ is designed such that the information signal sent from the PU to the SU about the  primary queue current state, empty or nonempty, is received correctly with probability one at the SU. This is important for designing an efficient access protocol on the basis of the actual state of the time slot, i.e., busy/free. As mentioned earlier, nodes dedicate a small band to cooperatively exchange information regarding the actual state of the PU. The transmission of the state occurs over the time interval $[0,\tau]$, where $\tau$ is assumed to be the transmission time of the information and is chosen to result in a negligible decoding errors of these information at the SU.

              We summarize the medium access control (MAC) as follows:

                            \begin{itemize}

\item The PU transmits the packet at the head of its queue.
\item During the time interval $[0,\tau]$, the PU sends its queue state (empty or nonempty) to the SU over the dedicated bandwidth for information exchange.
\item If the SU has energy packets and decides to access the channel at the beginning of the time slot, it ignores the information sent from the PU and resumes its transmission till the end of the time slot. This happens with probability $1-f$.
    \item If at the beginning of the time slot the SU decides to receive the primary packet, which happens with probability $f$, it adjusts its receiving end to the receiving mode and starts to collect data from the primary transmission.
         \item Based on the received state signal from the PU, the SU perfectly discerns the state of the PU.
              \item If the PU's queue is nonempty and the secondary energy
queue is nonempty, the SU decides whether to resume primary packet reception, which occurs with probability $\omega$; or to access the channel concurrently with the PU using one of its data queues, which occurs with probability $\overline{\omega}$. In the latter case, accessing the channel simultaneously with the PU is motivated by the presence of the MPR capability at receivers.
                  \item If the PU's queue is empty and the secondary energy queue is nonempty, the SU accesses the channel with probability $1$ using one of its data queues.
\item If at the beginning of the time slot the SU has no energy packets in its energy queue, it decides whether to receive the primary packet, which occurs with probability $\alpha$, or not.\footnote{We assume very small energy needed for packets decoding, which is reasonable due to its small value relative to the energy per packet (or energy needed for data transmission).} Note that since there is no energy in the secondary energy queue, there is no need to take another decision at $t\!=\!\tau$ seconds. This is because the SU is incapable of establishing any data transmission due to the lack of energy. In such cases, the probability of receiving the primary packet is $\alpha$, whereas the probability of remaining silent till the end of the current time slot is $\overline{\alpha}$.

\item At the far end of the time slot, the SU decides, on the basis of its ability to decode the primary packet and the status of primary packet decoding at the primary destination, whether to accept or reject the admission of the primary packet to the relaying queue. The acceptance probability of a primary packet is $\beta$, whereas the rejection probability is $\overline{\beta}\!=\!1\!-\!\beta$.
\end{itemize}
If the relaying queue is nonempty, the SU selects one of its packets for transmission with probability $\overline{\Gamma}\!=\!1-\Gamma$; or selects one of the relaying packets with probability $\Gamma$. If the relaying queue is empty, the SU accesses the channel using its own packets with probability $1$. The selection probability $\Gamma$ represents the
relative importance of the primary relaying packets and is used for controlling the throughput of the relaying queue. Choosing $\Gamma \!=\!1$ gives full priority to the relaying packets over the secondary packets, while $\Gamma\!=\!0$ favors the
secondary packets (i.e., no selection for the relaying packets). By varying $\Gamma$ between
$0$ and $1$, we can maximize the secondary throughput under stability of the other queues.

We would like to emphasize here the importance of having different parameters associated with the different state of the queues in the system. Having such parameters enhance the system performance and help in achieving the optimal performance of the network under investigation.

It should be noted that the probability of outage of a certain link depends on the time available for data transmission. Hence, the probability of outage when the SU transmits at the beginning of the time slot is less than the probability of outage when it starts data transmission at $t\!=\!\tau$. Although using lower transmission time raises the secondary transmitted power, ${\rm e}/(T-\tau)$, the channel outage raises as well \cite{wimob,ElSh1312:Optimal} (see Appendix~A for proof). We should note that the interference caused by the SU on the PU's transmission increases with the delay in secondary data transmission. This happens because the secondary transmit power raises as mentioned earlier. The reader is referred to Appendix A for more details.

At the far end of each time slot, a feedback acknowledgement/negative-acknowledgement (ACK/NACK) signal is sent from the receiver to inform the respective transmitter about the decodability status of its packet. The feedback message is overheard by all nodes in the network due to the wireless channel broadcast nature. Decoding errors of the feedback messages at the transmitters are negligible, which is reasonable for short length packets as low rate and strong codes can be employed in the feedback channel \cite{krikidis2010stability,sadek}. If a packet is received correctly at its destination, it is then removed from the system.

For the primary packets, if the primary destination can decode the transmitted packet, it sends back an ACK and the packet leaves the system. If the SU can decode the packet and the packet is admitted (accepted) for relaying while the primary destination cannot, the SU sends back an ACK and the PU drops that packet. If the SU cannot decode the primary packet; or if it can correctly decode the packet but decides to reject it and the primary destination fails in decoding the packet, the PU retransmits that packet at the following time slot. We note that the feedback signals sent by the SU and the primary destination are separated either in time or frequency.
\subsection{Queues Service and Arrival Processes}
 Let us first consider the packets of the primary queue, $Q_{\rm p}$. A packet departs the primary queue in either one of the following events. If the link ${\rm p\rightarrow d_p}$ is not in outage; or if the link ${\rm p\rightarrow d_p}$ is in outage, the link ${\rm p\rightarrow s}$ is not in outage, and the SU decides to admit the packet to the relaying queue. A successfully received packet by either the primary destination or the SU will be dropped from the primary queue. The mean service rate of the primary queue is then given by
\begin{equation}
\begin{split}
\mu_{\rm p}\!&=\! \overline{P_{\rm pd_p,0}} \Biggr(\!\overline{{\rm Pr}\{Q_{\rm e}\!\ne\! 0\}}\!+\!f{\rm Pr}\{Q_{\rm e}\ne 0\}\omega\!\\& \,\,\,\,\,\,\,\,\,\,\,\,\,\,\,\,\,\,\,\,\,\,\,\,\,\,\,\,\,\,\,\ +\!  {\rm Pr}\{Q_{\rm e}\ne 0\}(\delta_{\rm pd_p,00}\overline{f}+\delta_{\rm pd_p,01} f\overline{\omega})\Biggr)\\& \!+\!  P_{\rm pd_p,0} \overline{P_{\rm ps,0}} (\alpha\overline{{\rm Pr}\{Q_{\rm e}\ne 0\}}+  f {\rm Pr}\{Q_{\rm e}\ne 0\}\omega) \beta
\label{mup}
\end{split}
\end{equation}
where $\delta_{\rm pd_p,00}$ and $\delta_{\rm pd_p,01}$ denote the reduction in $\overline{P_{\rm pd_p,0}}$ due to concurrent transmission when the SU accesses the channel at $t\!=\!0$ and $t\!=\!\tau$, respectively. The definition and derivation of $\overline{P_{jk,i}}$ and $\delta_{jk,in}$ are provided in Appendix A. It should be pointed out here that without cooperation the maximum mean service rate for the primary queue is $\overline{P_{\rm pd_p,0}}$, whereas with cooperation the maximum achievable primary mean service rate is $\overline{P_{\rm pd_p,0}}~+~ P_{\rm pd_p,0}\overline{P_{\rm ps,0}}$, which is attained when the SU sets $\beta\!=\!\alpha\!=\!f\!=\!\omega\!=\!1$. Thus, the maximum achievable throughput of the PU is increased by $ P_{\rm pd_p,0}\overline{P_{\rm ps,0}}$ packets per time slot.

A packet from $Q_{\rm s}$ is served if the secondary energy queue is nonempty, the SU decides to access the channel using $Q_{\rm s}$, and the link ${\rm s\rightarrow d_s}$ is not in outage. The mean service rate of $Q_{\rm s}$ is given by
\begin{equation}
\begin{split}
\mu_{\rm s}&\!=\! \overline{P_{\rm sd_s,0}}\Big(\!\overline{f} \Big(\!{\rm Pr}\{Q_{\rm p}\!\ne\! 0,Q_{\rm e}\!\ne\! 0\}\delta_{\rm sd_s,00} \!+\!{\rm Pr}\{Q_{\rm p}\!=\! 0,Q_{\rm e}\!\ne\! 0\}\!\Big)\\& \!\,\,\,\,\,\,\,\,\,\,\,\,\,\,\,\,\,\,\,\,\,\ +\!\hat{\delta}_{\rm sd_s}f\Big(\!\overline{\omega}{\rm Pr}\{Q_{\rm p}\!\ne\! 0,Q_{\rm e}\!\ne\! 0\}\delta_{\rm sd_s,10}\!\\& \,\,\,\,\,\,\,\,\,\,\,\,\,\,\,\,\,\,\,\,\,\,\,\,\,\,\,\,\,\,\,\,\,\,\,\,\,\,\,\,\,\,\,\,\,\,\,\,\,\,\,\,\,\,\,\,\,\,\,\,\,\,\,\,\,\,\,\,\,\,\,\,\ +\!{\rm Pr}\{Q_{\rm p}\!=\! 0,Q_{\rm e}\!\ne\! 0\}\!\Big)\!\Big)\\& \,\,\,\,\,\,\,\,\,\,\,\,\,\,\,\,\,\,\,\,\,\,\,\,\,\,\,\,\,\,\,\,\,\,\,\,\,\,\,\,\,\,\,\,\,\,\,\,\,\,\,\,\,\,\,\,\,\,\,\ \times  \Big(\overline{\Gamma} {\rm Pr}\{Q_{\rm r}\!\ne\! 0\}+\overline{{\rm Pr}\{Q_{\rm r}\!\ne\! 0\}}\Big),
\end{split}
\end{equation}
where $\hat \delta_{jk}\!=\!\frac{\overline{P_{jk,1}}}{\overline{P_{jk,0}}}$ is defined in Appendix A.

Similarly, the mean service rate of $Q_{\rm r}$ is given by
\begin{equation}
\begin{split}
\mu_{\rm r}&\!=\! \overline{P_{\rm sd_p,0}}\Gamma\Big(\!\overline{f} \Big(\!{\rm Pr}\{\!Q_{\rm p}\!\ne\! 0,\!Q_{\rm e}\!\ne\! 0\!\}\delta_{\rm sd_p,00}\!+\!{\rm Pr}\{\!Q_{\rm p}\!=\! 0,\!Q_{\rm e}\!\ne\! 0\!\}\!\Big)\\& \!\,\,\,\,\,\,\,\,\,\,\,\,\,\,\,\,\,\,\,\,\,\,\,\,\,\,\,\,\,\,\,\ +\!\hat{\delta}_{\rm sd_p}f\Big(\!\overline{\omega}{\rm Pr}\{Q_{\rm p}\!\ne\! 0,Q_{\rm e}\!\ne\! 0\}\delta_{\rm sd_p,10}\!\\& \,\,\,\,\,\,\,\,\,\,\,\,\,\,\,\,\,\,\,\,\,\,\,\,\,\,\,\,\,\,\,\,\,\,\,\,\,\,\,\,\,\,\,\,\,\,\,\,\,\,\,\,\,\,\,\,\,\,\,\,\,\,\,\,\,\,\,\,\,\,\,\,\,\,\,\,\,\,\,\,\,\,\,\ +\!{\rm Pr}\{Q_{\rm p}\!=\! 0,Q_{\rm e}\!\ne\! 0\}\!\Big)\!\Big).
\end{split}
\end{equation}

The mean arrival rate of the relaying queue is obtained directly from (\ref{mup}). That is,
\begin{equation}
\begin{split}
\lambda_{\rm r}\!=\! P_{\rm pd_p,0} \overline{P_{\rm ps,0}} \Bigg(\!\alpha\overline{{\rm Pr}\{Q_{\rm e}\ne 0\}}+  f {\rm Pr}\{Q_{\rm e}\!\ne\! 0\}\omega\!\Bigg) \beta{\rm Pr}\{Q_{\rm p}\!\ne\! 0\},
\label{lambdar}
\end{split}
\end{equation}
where ${\rm Pr}\{Q_{\rm p}\ne 0\}$ in (\ref{lambdar}) means that the arrival of a primary packet at $Q_{\rm r}$ occurs when the primary queue is nonempty.

An energy packet is consumed from the secondary energy queue in a time slot if the SU decides to transmit a data packet from one of its data queues. The mean service rate of $Q_{\rm e}$ is then given by
\begin{equation}
\begin{split}
\mu_{\rm e}\!&\!=\!\!\overline{f}\!+\!{\rm Pr}\{Q_{\rm p}\ne 0\}f\overline{\omega}+f \overline{{\rm Pr}\{Q_{\rm p}\!\ne\! 0\}}\!=\!1\!-\!{\rm Pr}\{Q_{\rm p}\ne 0\}f\omega.
\label{goog}
\end{split}
\end{equation}
In (\ref{goog}), $\overline{f}$ means that the SU accesses the channel at $t\!=\!0$; ${\rm Pr}\{Q_{\rm p}\ne 0\}f\overline{\omega}$ means that the SU decides to access the channel at $t\!=\!\tau$ seconds, which occurs with probability $\overline{\omega}$ when $\{Q_{\rm p}\ne 0\}$; and $f \overline{{\rm Pr}\{Q_{\rm p}\!\ne\! 0\}}$ means that the SU decides to access the channel after $\tau$ seconds with probability one when $\{Q_{\rm p}\!=\! 0\}$.

Relaying the primary packets by the SU may seem to waste the time slots that could be used otherwise for its own packets. However, it
turns out that the SU is indeed gaining since
opportunistic relaying of primary packets results in emptying (servicing)
the primary queue faster as the service process of the primary queue increases; in return, more network resources can
be utilized for delivering the secondary packets. As a
result, all users simultaneously achieve performance gains.
\subsection{Approximated Systems}
The service processes of the primary data queue and the secondary energy queue are coupled, i.e., interacting queues. This means that the departure of a packet at any of them depends on the state of the other. Hence, we cannot analyze the system performance or compute the service process of each queue directly. For this reason, we study three approximated systems. Two of them provide inner bounds and the third provides an outer bound on the actual performance.

In the first approximated system, we assume that the PU transmits dummy packets when its queue is empty. These packets may interfere with the SU in case of concurrent transmissions, but do not contribute on the throughput of the PU. The essence of such assumption is to cause a constant interference with the SU to decouple the queue interaction and to render the computation of nodes' service rates possible. Under such assumption, the probability of the primary queue being empty is set to zero; that is, ${\rm Pr}\{Q_{\rm p}=0\}=0$ and ${\rm Pr}\{Q_{\rm p}\ne0\}=1$.\footnote{This is actually the stochastic dominance approach extensively investigated in the literature, see for example \cite{rao1988stability,luo1999stability,krikidis2012stability,sadek,wimob,ourletter}.} Since the PU is always backlogged (has at least one packet at its queue in each time slot), the probability of the SU finds a free time slot is zero. Thus, all time slots that the SU decides to access in are occupied by the PU. Hence, the service rates of the secondary queues, $Q_{\rm s}$ and $Q_{\rm r}$, are reduced relative to the original system in which the PU's queue may be empty in some time slots and the SU can access the channel alone.\footnote{Accessing the channel alone (without interference) provides a successful packet decoding at the relevant receiver higher than the case of concurrent transmission as is obvious. The reader is referred to Appendix A for proofs and further details.} Accordingly, this system is an inner bound for the original system.

In the second approximated system, we assume an energy packet dissipation in each time slot, which implies that $\mu_{\rm e}\!=\!1$ energy packet per time slot. Under such assumption, the probability of the energy queue being empty is significantly increased relative to the original system.\footnote{This is actually the highest probability for a queue to be empty because the service rate is $1$ packets/slot.} Consequently, the secondary packets get service less frequently. Furthermore, the relaying packets get service in a lower rate, hence the event of primary queue being empty decreases as the SU may decrease the acceptance ratio of the relaying packets to maintain its relaying queue stability. Thus, the possibility of having a free time slot or an interference-free time slot for the SU is reduced as well. Accordingly, this system is an inner bound on the original system.

In the third approximated system, we assume that the departure of the energy queue is almost zero, or equivalently, the probability of having an energy packet stored in the secondary energy queue in any time slot is one. This system is an outer bound on the original system as the SU will always be able to access the channel for transmitting its own packets or retransmitting the relayed primary packets each time slot, if there is a chance for the SU to access the channel. Hence, all service rates of the data queues will be increased simultaneously.
\subsubsection{First approximated system, Inner bound}
In this case, denoted by $\mathcal{S}_1$, the PU is always backlogged. If the SU decides not to access the channel at the beginning of the time slot, it will not access later at $t\!=\!\tau$. This is because the PU is always active and wasting $\tau$ seconds for knowing the activity state of the PU will not lead to any gains in terms of secondary queues throughput. Therefore, the optimal $\omega$ is $\omega^*=1$. Moreover, the decision on accessing the channel or receiving of the primary packet is taken at the early beginning of the time slot, specifically at $t=0$. If the secondary energy queue is nonempty, the SU decides to access the channel by one of its queue with probability $\overline{f}$ or decides to receive the possible primary transmission with probability $f$. If the secondary energy queue is empty, the SU cannot transmit data and its decision becomes whether to receive of the possible primary transmission with probability $\alpha$ or remain idle with probability $1\!-\!\alpha$. At the end of the time slot, the SU decides whether to admit the primary packet or to reject it, as explained earlier. Under the first approximated system, the mean service rate of the energy queue is given by
\begin{equation}
\begin{split}
\mu_{\rm e}\!&=\!1-f.
\end{split}
\end{equation}

Using the results provided in Appendix B (setting $\mu\!=\!\mu_{\rm e}\!=\!1\!-\!f$), the probability of the energy queue being empty is given by
\begin{equation}
\nu_{\circ}\!=\!1-\frac{\lambda_{\rm e}}{1-f}.
\end{equation}

Based on this, the relaying queue departure and arrival mean rates are, respectively, given by
\begin{equation}
\begin{split}
\mu_{\rm r}\!=\! \overline{P_{\rm sd_p,0}} \Gamma \overline{f} \overline{\nu_\circ}\delta_{\rm sd_p,0},
\label{mur1}
\end{split}
\end{equation}
\begin{equation}
\begin{split}
\lambda_{\rm r}\!=\!  P_{\rm pd_p,0} \overline{P_{\rm ps,0}} \Bigg(\alpha\nu_\circ\!+\!f \overline{\nu_\circ}\Bigg)  \beta.
\label{lambdar1}
\end{split}
\end{equation}
The probability of the relaying queue being nonempty is given by\footnote{The expression in (\ref{pir}) is obtained via solving the Markov chain modeling the relaying queue when its arrival and service processes are decouple of the other queue and become computable.}
\begin{equation}
\begin{split}
{\rm Pr}\{Q_{\rm r}\ne0\}\!=\!\pi_{\rm r}\!=\!\frac{\lambda_{\rm r}}{\mu_{\rm r}}.
\label{pir}
\end{split}
\end{equation}
The mean service rate of $Q_{\rm s}$ becomes
\begin{equation}
\begin{split}
\mu_{\rm s}\!=\! \overline{P_{\rm sd_s,0}}\ \overline{f} \overline{\nu_\circ}\delta_{\rm sd_s,00} \Big(\overline{\Gamma} \pi_{\rm r}+\overline{\pi_{\rm r}}\Big).
\label{mus1}
\end{split}
\end{equation}

The primary queue mean service rate is given by
\begin{equation}
\begin{split}
\mu_{\rm p}\!=&\! \overline{P_{\rm pd_p,0}} \Big((1\!-\!\overline{\nu_\circ}\ \overline{f})\! +\! \delta_{\rm pd_p,00} \overline{f}\overline{\nu_\circ}\Big)\!\\& \,\,\,\,\,\,\,\,\,\,\,\,\,\,\,\,\,\,\,\,\,\,\,\,\,\ +\!  P_{\rm pd_p,0} \overline{P_{\rm ps,0}} (\alpha\nu_\circ\!+\!f \overline{\nu_\circ})  \beta.
\label{mup1}
\end{split}
\end{equation}

We note that the queues are not interacting anymore. Hence, we can apply Loynes theorem to check the stability of the queues and obtain the maximum stable throughput based on the first approximated system via solving the following constrained optimization problem.

\begin{equation}
\begin{split}
     &\underset{\beta,f,\alpha,\Gamma}{\max.}\,\,\,\,\,\,\,\,\,\,\,\ \mu_{\rm s}, \,\,\ {\rm s.t.} \,\,\,\,\,\,\,\   \lambda_{\rm r}\le \mu_{\rm r}, \lambda_{\rm p}\le \mu_{\rm p},
     \label{opt2}
     \end{split}
    \end{equation}
where  $\mu_{\rm r}$, $\lambda_{\rm r}$, $\mu_{\rm s}$ and $\mu_{\rm p}$ are in (\ref{mur1}), (\ref{lambdar1}), (\ref{mus1}) and (\ref{mup1}), respectively.

For a given $f$ and $\beta$, we can get a closed-form expressions for $\Gamma$ and $\alpha$, then we solve a family of convex optimization problems parameterized by $\beta$ and $f$. Specifically, the optimal solutions of $\Gamma$ and $\alpha$ are a set of points which satisfies the stability constraint of the primary and relaying queues stability, respectively. Using (\ref{mup1}), the optimal $\alpha$ for a fixed $f$ and $\beta$ is given by
\begin{equation}
\begin{split}
 \alpha^*\!\ge\!\frac{\frac{\lambda_{\rm p}\!-\! \overline{P_{\rm pd_p,0}} \Big((1\!-\!\overline{\nu_\circ}\ \overline{f})\! +\! \delta_{\rm pd_p,00} \overline{f}\overline{\nu_\circ}\Big)}{ P_{\rm pd_p,0} \overline{P_{\rm ps,0}}\beta}\!-\!f \overline{\nu_\circ}}{\nu_\circ}.
 \label{alpha}
\end{split}
\end{equation}

Using the constraint on the stability of the relaying queue, the optimal $\Gamma$ is given by
 \begin{equation}
\begin{split}
\Gamma^*\ge\frac{  P_{\rm pd_p,0} \overline{P_{\rm ps,0}} (\alpha^*\nu_\circ\!+\!f \overline{\nu_\circ})  \beta}{\overline{P_{\rm sd_p,0}}\ \overline{f} \overline{\nu_\circ}\delta_{\rm sd_p,0}},
\label{gamma}
\end{split}
\end{equation}
where $\alpha^*$ is given in (\ref{alpha}). The optimal $\beta$ and $f$ are obtained via grid search and are selected as the pair of parameters that yields the highest objective function in (\ref{opt2}). From (\ref{gamma}), we note that the optimal selection probability of the relaying queue for transmission, $\Gamma^*$, increases with increasing the acceptance probability of the primary undelivered packets, $\beta$, and the flow rate to the relaying queue ${  P_{\rm pd_p,0} \overline{P_{\rm ps,0}} (\alpha^*\nu_\circ\!+\!f \overline{\nu_\circ})  \beta}$. This is because the SU should increase the selection of $Q_{\rm r}$ for transmission to maintain the relaying queue stability. In addition, $\Gamma^*$ increases with decreasing of $\overline{P_{\rm sd_p,0}}\ \overline{f} \overline{\nu_\circ}\delta_{\rm sd_p,0}$. This is because $\overline{P_{\rm sd_p,0}}\ \overline{f} \overline{\nu_\circ}\delta_{\rm sd_p,0}$ determines the probability of certain transmitted packet from the relaying queue being correctly received at the primary destination and therefore if this term is high, the SU will not need several transmission for the same packet each time slot. Hence, the SU can reduce the probability of choosing the relaying queue for transmission at a time slot and rather it could use that time slot for the transmission of its own packets.
\subsubsection{Second approximated system, Inner bound}
In this approximated system, denoted by $\mathcal{S}_2$, we assume that an energy packet is consumed per time slot. That is, $\mu_{\rm e}\!=\!1$ energy packets per time slot. The probability of the energy queue being empty is given by
\begin{equation}
\begin{split}
\nu_{\circ}\!=\! 1\!-\!\frac{\lambda_{\rm e}}{\mu_{\rm e}}\!=\!1\!-\!\lambda_{\rm e}.
\end{split}
\end{equation}
We can interpret the probability $\lambda_{\rm e}$ as the fraction of time slots that can be used by the SU for data transmission. It should be pointed out here that under this approximation the buffer size does not change the state probabilities.
Hence, does not have any impact on the queues' rates. The Markov chain modeling the energy queue in this case is composing of two states only: state $0$ where the energy queue has no packets, and state $1$ where the energy queue has only one packet. The probability of the energy queue having more than one packet, $\nu_k$, $k\ge2$, is zero.

 The mean service rate of $Q_{\rm p}$ is given by
 \begin{equation}
\begin{split}
\mu_{\rm p}\!&=\! \overline{P_{\rm pd_p,0}} \Big((\overline{\lambda_{\rm e}}+f\lambda_{\rm e}\omega)\!+\!  \lambda_{\rm e} (\delta_{\rm pd_p,00}\overline{f}+\delta_{\rm pd_p,01} f\overline{\omega})\Big)\!\\&  \,\,\,\,\,\,\ \,\,\,\,\,\,\,\,\,\,\,\,\,\,\,\,\,\,\,\,\,\,\,\,\,\ +\!  P_{\rm pd_p,0} \overline{P_{\rm ps,0}} (\alpha \overline{\lambda_{\rm e}}+ f\lambda_{\rm e}\omega) \beta.
\label{mup2}
\end{split}
\end{equation}
The probability of the primary queue being nonempty is given by

\begin{equation}
\begin{split}
{\rm Pr}\{Q_{\rm p}\ne0\}\!=\!\pi_{\rm p}\!=\!\frac{\lambda_{\rm p}}{\mu_{\rm p}}.
\label{pip}
\end{split}
\end{equation}
The relaying queue mean service and arrival rates are, respectively, given by

\begin{equation}
\begin{split}
\mu_{\rm r}\!=\! \lambda_{\rm e} \overline{P_{\rm sd_p,0}} \Gamma \Big(\overline{f}\Big(\!\pi_{\rm p}\delta_{\rm sd_p,00}\!+\!\overline{\pi_{\rm p}}\!\Big)\!+\!f\hat{\delta}_{\rm sd_p}\Big(\!\delta_{\rm sd_p,10}\overline{\omega}\pi_{\rm p}\!+\!\overline{\pi_{\rm p}}\!\Big)\!\Big),
\label{mur2}
\end{split}
\end{equation}

\begin{equation}
\begin{split}
\lambda_{\rm r}\!=\! P_{\rm pd_p,0} \overline{P_{\rm ps,0}} \Bigg(\alpha \overline{\lambda_{\rm e}}\!+\! f\lambda_{\rm e}\omega\Bigg) \beta\pi_{\rm p}.
\label{lambdar2}
\end{split}
\end{equation}

The probability of the relaying queue being nonempty is given by\footnote{The expression in (\ref{pir}) is obtained via solving the Markov chain modeling the relaying queue when its arrival and service processes are decouple of the other queue and become computable.}
\begin{equation}
\begin{split}
{\rm Pr}\{Q_{\rm r}\ne0\}\!=\!\pi_{\rm r}\!=\!\frac{\lambda_{\rm r}}{\mu_{\rm r}}.
\label{pir}
\end{split}
\end{equation}

The mean service rate of $Q_{\rm s}$ is then given by
\begin{equation}
\begin{split}
\mu_{\rm s}&\!=\! \overline{P_{\rm sd_s,0}}\lambda_{\rm e}\Big(\overline{f} \Big(\!\pi_{\rm p}\delta_{\rm sd_s,00}\!+\!\overline{\pi_{\rm p}}\!\Big)\!+\!\hat{\delta}_{\rm sd_s}f\Big(\!\overline{\omega}\pi_{\rm p}\delta_{\rm sd_s,10}\!+\!\overline{\pi_{\rm p}}\!\Big)\!\Big)\\&\,\,\,\,\,\,\,\,\,\,\,\,\,\,\,\,\,\,\,\,\,\,\,\,\,\,\,\,\,\,\,\,\,\,\,\,\,\,\,\,\,\,\,\,\,\,\,\,\,\,\,\ \times \Big(\overline{\Gamma} \pi_{\rm r}\!+\!\overline{\pi_{\rm r}}\Big).
\label{mus2}
\end{split}
\end{equation}
Since the queues are decoupled in the second approximated system, the maximum secondary throughput is given by solving the following problem.
\begin{equation}
\begin{split}
     &\underset{\beta,f,\alpha,\omega,\Gamma}{\max.}\,\,\,\,\,\,\,\,\,\,\,\ \mu_{\rm s}, \,\,\ {\rm s.t.} \,\,\,\,\,\,\,\   \lambda_{\rm r}\le \mu_{\rm r}, \lambda_{\rm p}\le \mu_{\rm p},
     \label{optim2}
     \end{split}
    \end{equation}
where $\mu_{\rm p}$, $\mu_{\rm r}$, $\lambda_{\rm r}$ and $\mu_{\rm s}$ are in (\ref{mup2}), (\ref{mur2}), (\ref{lambdar2}), and (\ref{mus2}), respectively.

We conjecture that the throughput region of the second approximated system contains that of the first approximated system. This is because, in contrast to the second approximated system where there can be free time slots, in the first approximated system, the PU is always backlogged; hence, it always interferes with the SU and the probability of a free time slot for the SU is zero. Since the probability of success transmission under interference is low, the service rates of the SU's queues are decreased significantly under the first approximated system. Based on that, the second approximated system always provides better performance than the first approximated system.
\subsubsection{Third approximated system, Outer bound}
In this case, denoted by $\mathcal{S}_3$, we consider a backlogged energy queue. This means that there exists at least one energy packet each time slot in $Q_{\rm e}$. This case can happen
when $\lambda_{\rm e}\!=\!1$ energy packets/slot regardless of the value of $\mu_{\rm e}$. In this case, the probability of the energy queue being nonempty approaches the unity. The service and arrival rates are obtained directly from (1), (2), (3), (4) and (5) with $\Pr\{Q_{\rm e}\ne0\}=1$.
 The mean service and arrival rates of the queues are then given by
\begin{equation}
\begin{split}
\mu_{\rm p}\!&=\!\overline{P_{\rm pd_p,0}} \Big(f\omega\! +\!  (\delta_{\rm pd_p,00}\overline{f}+\delta_{\rm pd_p,01} f\overline{\omega})\Big)\!+\!  P_{\rm pd_p,0} \overline{P_{\rm ps,0}} f \omega \beta,
\label{mup3}
\end{split}
\end{equation}
\begin{equation}
\begin{split}
\mu_{\rm r}\!=\! \overline{P_{\rm sd_p,0}} \Gamma \Big(\!\overline{f}  \Big(\pi_{\rm p}\delta_{\rm sd_p,00}\!+\!\overline{\pi_{\rm p}}\Big)+f \hat{\delta}_{\rm sd_p} \Big(\overline{\omega}\pi_{\rm p}\delta_{\rm sd_p,10}\!+\!\overline{\pi_{\rm p}}\Big)\!\Big),
\label{mur3}
\end{split}
\end{equation}

\begin{equation}
\begin{split}
\lambda_{\rm r}\!=\! P_{\rm pd_p,0} \overline{P_{\rm ps,0}} f  \omega\beta\pi_{\rm p},
\label{lambdar3}
\end{split}
\end{equation}
where $\pi_{\rm p}$ in (\ref{mur3}) and (\ref{lambdar3}) follows (\ref{pip}) with $\mu_{\rm p}$ in (\ref{mup3}), and
\begin{equation}
\begin{split}
\mu_{\rm s}&\!=\! \overline{P_{\rm sd_s,0}}\Big(\overline{f}  \Big(\pi_{\rm p}\delta_{\rm sd_s,00}+\overline{\pi_{\rm p}}\Big)\!+\!f \hat{\delta}_{\rm sd_s}\Big(\overline{\omega}\pi_{\rm p}\delta_{\rm sd_s,10}\!+\!\overline{\pi_{\rm p}}\Big)\Big)\\& \,\,\,\,\,\,\,\,\,\,\,\,\,\,\,\,\,\,\,\,\,\,\,\,\,\,\,\,\,\,\,\,\,\,\,\,\,\,\,\,\,\,\,\,\,\,\,\,\,\,\,\,\,\,\,\,\,\,\,\,\,\,\,\,\,\ \times \Big(\overline{\Gamma} \pi_{\rm r}\!+\!\overline{\pi_{\rm r}}\Big),
\label{mus3}
\end{split}
\end{equation}
where $\pi_{\rm r}$ follows (\ref{pir}) with $\mu_{\rm r}$ and $\lambda_{\rm r}$ in (\ref{mur3}) and (\ref{lambdar3}), respectively.

The outer bound can be computed by solving the following problem:
\begin{equation}
\begin{split}
     &\underset{\beta,f,\omega,\Gamma}{\max.}\,\,\,\,\,\,\,\,\,\,\,\ \mu_{\rm s}, \,\,\ {\rm s.t.} \,\,\,\,\,\,\,\   \lambda_{\rm r}\le \mu_{\rm r}, \lambda_{\rm p}\le \mu_{\rm p},
     \label{optim3}
     \end{split}
    \end{equation}
    where $\mu_{\rm p}$, $\mu_{\rm r}$, $\lambda_{\rm r}$ and $\mu_{\rm s}$ are in (\ref{mup3}), (\ref{mur3}), (\ref{lambdar3}), and (\ref{mus3}), respectively.

 The optimization problems are solved numerically at the SU for a given channels and system parameters. Specifically, for a given parameters, the SU solves the optimization problem and use the optimal parameters for the system's operation.

\subsection{Some Important Remarks}
Following are some important remarks.
\subsubsection{First Remark} Using the results in Appendix A, the complement of outage probability of link ${\rm p\rightarrow d_p}$ when the SU starts transmission at the beginning of the time slot is given by
 \begin{eqnarray}
 \overline{P_{\rm pd_p,00}^{\left({\rm c}\right)}}\!=\!\frac{1}{1\!+\!\Big( {2^{\frac{\mathcal{B}}{WT}}\!-\!1} \Big)\frac{\gamma_{\rm sd_p,0}\sigma_{\rm sd_p}}{ \gamma_{jk}\sigma_{jk}}} {\exp\Big(-\!\frac{{2^{\frac{\mathcal{B}}{WT}}\!-\!1}}{\gamma_{\rm pd_p,0} \sigma_{\rm pd_p}}\Big)},
 \label{df1}
\end{eqnarray}
while the probability of that link being not in outage when the SU starts transmission at $t\!=\!\tau$ is given by
 \begin{eqnarray}
  \label{dfx2}
  \overline{P_{\rm pd_p,01}^{\left({\rm c}\right)}}\!=\!\frac{1}{1\!+\!\Big( {2^{\frac{\mathcal{B}}{WT}}\!-\!1} \Big)\frac{\gamma_{\rm sd_p,1}\sigma_{\rm sd_p}}{ \gamma_{jk}\sigma_{jk}}} {\exp\Big(-\!\frac{{2^{\frac{\mathcal{B}}{WT}}\!-\!1}}{\gamma_{\rm pd_p,0} \sigma_{\rm pd_p}}\Big)}.
\end{eqnarray}
The ratio of (\ref{df1}) to (\ref{dfx2}) is given by
 \begin{eqnarray*}\label{193}
 \rho\!=\!\frac{ \overline{P_{\rm pd_p,01}^{\left({\rm c}\right)}}}{ \overline{P_{\rm pd_p,00}^{\left({\rm c}\right)}}}=\frac{{1+\Big( {2^{\frac{\mathcal{B}}{WT}}-1} \Big)\frac{\gamma_{\rm sd_p,0}\sigma_{\rm sd_p}}{ \gamma_{\rm pd_p,0}\sigma_{\rm pd_p}}} }{{1+\Big( {2^{\frac{\mathcal{B}}{WT}}-1} \Big)\frac{\gamma_{\rm sd_p,1}\sigma_{\rm sd_p}}{ \gamma_{\rm pd_p,0}\sigma_{\rm pd_p}}}} =\frac{1+a}{1+\frac{a}{1-\tau/T}}.
\end{eqnarray*}
We note that $\gamma_{\rm sd_p,1}\!=\!\gamma_{\rm sd_p,0}/(1\!-\!\tau/T)$ and $a\!=\!\Big( {2^{\frac{\mathcal{B}}{WT}}-1} \Big)\frac{\gamma_{\rm sd_p,0}\sigma_{\rm sd_p}}{ \gamma_{\rm pd_p,0}\sigma_{\rm pd_p}}$. If $a\gg1$, the reduction of the primary packet correct reception probability due to secondary access delay (when the SU accesses the channel at $t\!=\!\tau$) is $\rho\!\approx\!1\!-\!\tau/T$. Therefore, if the secondary decides to access after $\tau$ seconds of primary packet reception based on the primary activity, the probability of the primary packet decoding reduces by a factor $1-\tau/T$ relative to the case when the SU accesses the channel at the beginning of the slot. The reduction of the primary packet correct reception is a linear function of $\tau$. If the decision time, $\tau$, is high, the primary packet decoding will be reduced significantly.

Assume that the primary transmits with a very low power. This makes $a$ much greater than $1$. Thus, we can approximate the reduction, due to secondary access delay, of the probability of the primary channel not being in outage by $\rho\!\approx\!1-\tau/T$. At the same time, since the primary transmit power is low, the required $\tau$ for perfect primary detection is high. This means that the reduction of the primary packet decoding at the primary destination due to concurrent transmissions is significantly high. In this case, the secondary access probability at $t=0$ is definitely higher than the access probability at $t=\tau$ when the PU is detected to be active and the SU decides to accesses the channel. Moreover, it may be better for the SU to access the channel at $t=0$ to use the whole slot time in data transmission; and at $t=\tau$ if the PU is declared to be inactive, if the PU is declared to be active, it may be better to resume receiving the primary packet because concurrent transmission would be harmful for the PU as explained earlier.

\subsubsection{Second Remark}
Assume that the current primary arrival rate is $\lambda_{\rm p}=\lambda_{\rm p}^\star$. Increasing the primary arrival rate to $\lambda^\star_{\rm p}+\Delta_{\lambda_{\rm p}}$, $\Delta_{\lambda_{\rm p}}\ge0$, increases the probability of the primary queue being nonempty. This is because the probability of having an arrival at a certain time slot is increased. Consequently, the number of empty time slots that the SU can detect or access alone decreases as well. In addition, the probability of relaying queue selection, $\Gamma$, must be increased to maintain the stability of the relaying queue as the arrival rate of the relaying queue is increased due to the increasing of $\lambda_{\rm p}$. These two observations lead to the fact that the achievable secondary rate is increased relative to the case of $\lambda_{\rm p}=\lambda_{\rm p}^\star$. This means that the secondary service rate, $\mu_{\rm s}$, is a non-increasing function of the primary arrival rate $\lambda_{\rm p}$.

\subsubsection{Third Remark}
 From the expressions of the service rates of the queues, the service processes are functions of channel outages probabilities. Based on the formulas of the channel outage in Appendix~A, the outage probability of a certain link is a decreasing function of $\mathcal{R}_{\rm p}\!=\!\mathcal{B}/(TW)$. Therefore, increasing the targeted primary spectral efficiency rate, $\mathcal{R}_{\rm p}$, decreases all queues service rates. This leads to a reduction in the maximum achievable secondary throughput, $\mu_{\rm s}$. This means that the secondary service rate, $\mu_{\rm s}$, is a non-increasing function of the primary targeted spectral efficiency rate $\mathcal{R}_{\rm p}\!=\!\mathcal{B}/(TW)$.

The following proposition summarizes the main observations in the second and the third remarks.
\begin{proposition}
For a given channel and system parameters, let $\mu^*_{\rm s}(\lambda_{\rm p},R_{\rm p})$ be the maximum secondary throughput at the pair $(\lambda_{\rm p},R_{\rm p})$. The optimal secondary throughput satisfies the following properties:
\begin{itemize}
\item $\mu^*_{\rm s}(\lambda_{\rm p},R_{\rm p})\ge \mu^*_{\rm s}(\lambda_{\rm p}+\Delta_{\lambda_{\rm p}},R_{\rm p})$, $\Delta_{\lambda_{\rm p}}\ge0$.
\item $\mu^*_{\rm s}(\lambda_{\rm p},R_{\rm p})\ge \mu^*_{\rm s}(\lambda_{\rm p},R_{\rm p}+\Delta_{R_{\rm p}})$, $\Delta_{R_{\rm p}}\ge0$.
\end{itemize}
\end{proposition}
\section{Numerical Results and Conclusions}\label{numerical}
In this section, we provide some numerical results for the optimization problems presented in this paper. We define here the conventional scheme, denoted by $\mathcal{S}_{\rm c}$, where the SU senses the channel for $\tau$ seconds and if the primary data queue and the secondary energy queue are simultaneously empty and nonempty, respectively, the SU accesses the channel with probability $1$ using one of its queues probabilistically if the relaying queue is nonempty. In addition, if the PU is transmitting a packet to its destination, the SU accepts with probability one to relay and admit the transmitted packet if the primary destination fails in decoding that packet. The secondary throughput of the conventional system is obviously a subset of the proposed cooperative system, $\mathcal{S}$, and can be obtained from $\mathcal{S}$ via setting $\beta=1$, $\alpha\!=\!1$, $f=1$ and $\omega\!=\!0$. The other parameters are optimized over their domains to achieve the maximum secondary throughput.

Fig. \ref{fig3} represents the maximum secondary throughput of the approximated systems of system $\mathcal{S}$. The figures are generated using the following common parameters: $\overline{P_{\rm sd_p,0}}=0.8$, $\delta_{\rm sd_p,00}=0.3$, $\overline{P_{\rm sd_s,0}}=0.7$, $\overline{P_{\rm ps,0}}=0.8$, $\delta_{\rm sd_s,00}=0.3$, $\overline{P_{\rm pd_p,0}}=0$, $\overline{P^{\left(\rm s\right)}_{\rm pd_p,00}}=\overline{P^{\left(\rm s\right)}_{\rm pd_p,01}}\!=\!0$, $\hat{\delta}_{\rm sd_p}=0.7$, $\hat{\delta}_{\rm sd_s}=0.7$, $\delta_{\rm sd_p,10}=0.2$, $\delta_{\rm sd_s,10}=0.2$. The outer bound which represents the case of backlogged energy queue is close to the inner bound.

\begin{figure}
  \centering
  \includegraphics[width=1\columnwidth]{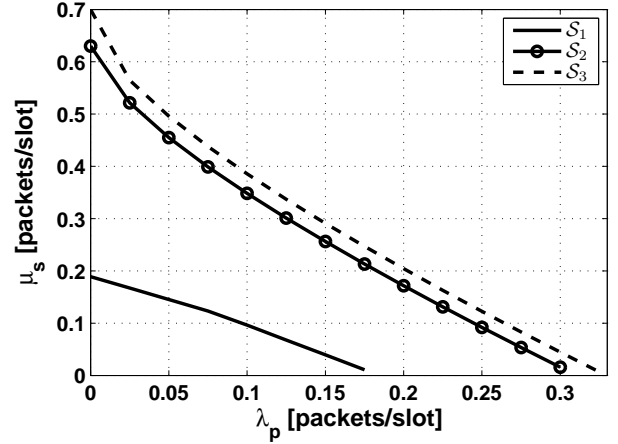}\\
  \caption{The maximum secondary throughput of the approximated systems for each $\lambda_{\rm p}$.}
  \label{fig2}
  \end{figure}

\begin{figure}
  \centering
  \includegraphics[width=1\columnwidth]{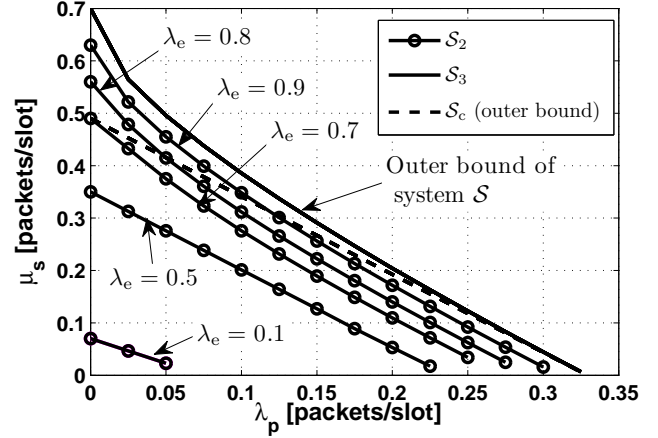}\\
  \caption{The maximum secondary throughput of the conventional cooperative protocol and the second and third approximated systems for each $\lambda_{\rm p}$ and for different values of $\lambda_{\rm e}$.}
  \label{fig3}
  \end{figure}

Figs. \ref{fig2} and \ref{fig3} represent the maximum secondary throughput of the approximated systems of system $\mathcal{S}$. The figures are generated using the following common parameters: $\overline{P{\rm sd_p,0}}=0.8$, $\delta_{\rm sd_p,00}=0.3$, $\overline{P_{\rm sd_s,0}}=0.7$, $\overline{P_{\rm ps,0}}=0.8$, $\delta_{\rm sd_s,00}=0.3$, $\mathcal{K}=60$, $\overline{P_{\rm pd_p,0}}=0$, $\overline{P^{\left(\rm s\right)}_{\rm pd_p,00}}=\overline{P^{\left(\rm s\right)}_{\rm pd_p,01}}\!=\!0$, $\hat{\delta}_{\rm sd_p}=0.7$, $\hat{\delta}_{\rm sd_s}=0.7$, $\delta_{\rm sd_p,10}=0.2$, $\delta_{\rm sd_s,10}=0.2$. In Fig. \ref{fig2}, the maximum secondary throughput under the approximated systems is plotted versus $\lambda_{\rm p}$. This figure is plotted with $\lambda_{\rm e}=0.9$ energy packets per time slot. The figure shows that the second approximated system provides throughput higher than the first approximated system, hence the union, which represents an inner bound on the actual performance of system $\mathcal{S}$, is the second approximated system. The outer bound, which represents the case of backlogged energy queue, is close to the inner bound. The gap between the two bounds shrinks as $\lambda_{\rm e}$ increases.

Fig. \ref{fig3} reveals two important observations. First, the figure reveals the impact of the arrival rate of the secondary energy queue on the system's inner bound.  Precisely, as the energy arrival rate increases, the inner and the outer bounds become close to each other and they overlap for all $\lambda_{\rm p}$ when $\lambda_{\rm e}\!=\!1$ energy packets/slot. Second, the figure reveals that the inner bound of the proposed system can outperform the outer bound of the conventional cooperation protocol with reliable energy source plugged to the SU. Note that system $\mathcal{S}_{\rm c}$ is plotted with $\lambda_{\rm e}\!=\!1$ energy packets per time slot (outer bound on $\mathcal{S}_{\rm c}$). We note that without cooperation the primary packets outage probability is $1\!-\!\overline{P_{\rm pd_p,0}}\!=\!1$ which implies that the probability of a primary packet being served in a given arbitrary time slot is zero. Hence, the primary queue is always backlogged and will never be empty. On the other hand, with cooperation the maximum feasible primary arrival rate is $0.3$ packets per time slot.

 We note that for Figs. \ref{fig2} and \ref{fig3}, without cooperation the primary packets outage probability is $1\!-\!\overline{P_{\rm pd_p,0}}\!=\!1$ which implies that the probability of a primary packet being served at an arbitrary time slot is zero. Hence, the primary queue is always backlogged and will never be emptied. On the other hand, with cooperation the maximum feasible primary arrival rate is $0.3$ packets per time slot.

\begin{figure}
  \centering
  \includegraphics[width=1\columnwidth]{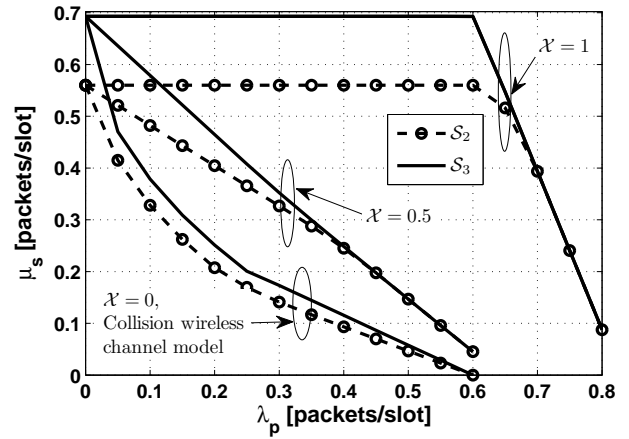}\\
  \caption{The maximum secondary throughput for each $\lambda_{\rm p}$ and for different values of MPR capability.
  }\label{fig5}
  \end{figure}

The impact of MPR capability is shown in Fig. \ref{fig5}. The figure reveals the gains of the MPR capability on achieving higher throughput for both users. The parameters are chosen to be: $\lambda_{\rm e}=0.8$,
$\overline{P_{\rm sd_p,0}}=0.8$, $\overline{P_{\rm sd_s,0}}=0.7$, $\overline{P_{\rm ps,0}}=0.8$, $\overline{P_{\rm pd_p,0}}=0.6$, $\hat{\delta}_{\rm sd_p}=0.5$, $\hat{\delta}_{\rm sd_s}=0.5$, and $\delta_{\rm pd_p,00}\!=\!\delta_{\rm sd_s,00}\!=\!\delta_{\rm sd_p,00}=\delta_{\rm sd_s,10}\!=\!\delta_{\rm sd_p,10}\!=\!\mathcal{X}$, which represents the MPR strength. At strong MPR, we can achieve orthogonal channels for terminals over most $\lambda_{\rm p}$ range. The plot also shows that the inner and the outer bounds coincide for high $\lambda_{\rm p}$. This happens because the energy queue is backlogged under the used parameters.

From the figures, it is noted that cooperation boosts both primary and secondary throughput. Furthermore, the energy arrival rate increases the probabilities of the secondary packets and the relayed primary packets being served which, in turn, boost both primary and secondary throughput. The figures also show that the increasing of $\lambda_{\rm p}$ decreases the maximum achievable secondary throughput.


\section{Conclusion}\label{conclusion}
In this paper, we have proposed a new cooperative cognitive relaying protocol, where the SU relays some of the undelivered primary packets. We have considered a generalized MPR channel model, and investigated the impact of the receivers' MPR capability on the users throughput. We also have investigated the impact of the secondary energy queue on the system performance. We have provided two inner bounds and an outer bound on the secondary throughput, and showed that the bounds are coincide when the secondary energy queue is always backlogged. The proposed protocol is designed such that the SU exploits the MPR capability and manages its energy packets to maximize its throughput under stability of the primary and the relaying queues.

A possible extension of this work can be directed to span the case of having an SU equipped with multiple antennas and with the availability of CSI at the transmitting antennas to achieve the maximum rates for the queues.


\section*{Appendix A}
We derive here a generic expression for the outage probability at the receiver of transmitter $j$ (node $k$) when there is a concurrent transmission from the transmitter $v$. Assume that node $j$ starts transmission at $i\tau$ and node $v$ starts transmission at $n\tau$. Outage occurs when the spectral efficiency  $R^{\left(i\right)}_j\!=\!\frac{\mathcal{B}}{WT^{\left(i\right)}_j}$ exceeds the channel capacity
\begin{equation} \small
P_{jk,in}^{\left({v}\right)}={\rm Pr}\biggr\{R^{\left(i\right)}_j >  \log_{2}\left(1+\frac{\gamma_{jk,i} \zeta_{jk}}{\gamma_{vk,n} \zeta_{vk}+1}\right)\biggr\}
\end{equation}
\noindent where ${\rm Pr}\{\mathcal{A}\}$ denotes the probability of the event $\mathcal{A}$, $\gamma_{jk,i}\!=\!\mathbb{P}^{\left(i\right)}_j/\mathcal{N}_k$, $\mathbb{P}^{\left(i\right)}_j$ is the used transmit power by node $j$ when it starts transmission at $t\!=\!i\tau$, $\gamma_{vk,n}\!=\!\mathbb{P}^{\left(n\right)}_v/\mathcal{N}_k$, and $\mathbb{P}^{\left(n\right)}_v$ is the used transmit power by node $v$ when it starts transmission at $t\!=\!n\tau$.
The outage probability can be written as
\begin{equation} \small\label {1900}
P_{jk,in}^{\left({v}\right)}={\rm Pr}\Big\{\frac{\gamma_{jk,i} \zeta_{jk}}{\gamma_{vk,n} \zeta_{vk}+1}<{2^{R^{\left(i\right)}_j}-1}\Big\}
\end{equation}
\noindent Since $\zeta_{jk}$ and $\zeta_{vk}$ are independent and exponentially distributed (Rayleigh fading channel gains) with means $\sigma_{jk}$ and $\sigma_{vk}$, respectively, we can use the probability density functions of these two random variables to obtain the outage as
 \begin{eqnarray}\label{193}
 P_{jk,in}^{\left({v}\right)}=1-\frac{1}{1+\Big( {2^{R^{\left(i\right)}_j}-1} \Big)\frac{\gamma_{vk,n}\sigma_{vk}}{ \gamma_{jk,i}\sigma_{jk}}} {\exp\Big(-\frac{{2^{R^{\left(i\right)}_j}-1}}{\gamma_{jk,i} \sigma_{jk}}\Big)}
\end{eqnarray}
We note that from the outage probability (\ref{193}), the numerator is increasing function of $R^{\left(i\right)}_j$ and the denominator is a decreasing function of $R^{\left(i\right)}_j$. Hence, the outage probability $P_{jk,in}^{\left({v}\right)}$ increases with $R^{\left(i\right)}_j$.
The probability of correct packet reception $\overline{P^{\left({v}\right)}_{jk,i}}=1-P^{\left({v}\right)}_{jk,i}$ is thus given by
  \begin{eqnarray}\label{conctra}
 \overline{P_{jk,in}^{\left({v}\right)}}=\frac{\overline{P_{jk,i}}}{1+\Big({2^{\frac{\mathcal{B}}{TW\left(1-\frac{i\tau}{T}\right)}}-1} \Big)\frac{\gamma_{vk,n}\sigma_{vk}}{\gamma_{jk,i} \sigma_{jk}}}\!=\!{\delta^{\left(v\right)}_{jk,in}}\overline{P_{jk,i}}
\end{eqnarray}
\noindent where $\overline{P_{jk,i}}\!=\!\exp\Big(-\frac{{2^{R^{\left(i\right)}_j}-1}}{\gamma_{jk,i} \sigma_{jk}}\Big)$ is the probability of correct packet reception when node $j$ transmits alone (without interference) and ${\delta^{\left(v\right)}_{jk,in}}\le1$ is the reduction in the probability of correct packet reception $\overline{P_{jk,i}}$ due to the presence of interference from node $v$. As is obvious, the probability of correct packet reception is lowered in the case of interference.
Based on (\ref{conctra}), we note that
  \begin{eqnarray}
 \frac{\overline{P_{jk,in}^{\left({v}\right)}}}{\overline{P_{jk,im}^{\left({v}\right)}}}=\!\frac{\delta^{\left(v\right)}_{jk,in}}{\delta^{\left(v\right)}_{jk,im}}
\end{eqnarray}
Following are some important notes. First, note that if the PU's queue is nonempty, the PU transmits its packet from the beginning of the time slot (at $t=0$) with a fixed transmit power $\mathbb{P}_{\rm p}$ and data transmission time $T_{\rm p}\!=\!T$. Accordingly, the superscript `$i$' in $T_j^{\left(i\right)}$ which represents the instant that a transmitting node starts its data transmission in is removed in case of PU. In addition, the superscript `$(v)$' is removed as we have only one PU and one SU.

Second, for the SU, the formula of probability of complement outage of link ${\rm s}\rightarrow k$ when the PU is active is given by
  \begin{eqnarray}\label{stconctra}
 \overline{P_{{\rm s} k,i0}^{\left({\rm p}\right)}}=\frac{\exp\Big(-\frac{{2^{\frac{\mathcal{B}}{TW\left(1-\frac{i\tau}{T}\right)}}-1}}{\gamma_{{\rm s} k,i} \sigma_{{\rm s} k}}\Big)}{1+\Big({2^{\frac{\mathcal{B}}{TW\left(1-\frac{i\tau}{T}\right)}}-1} \Big)\frac{\gamma_{{\rm p} k,0}\sigma_{{\rm p} k}}{\gamma_{{\rm s} k,i} \sigma_{{\rm s} k}}}\!
\end{eqnarray}
where $n\!=\!0$ because the PU always transmits at $t\!=\!0$ and $\gamma_{{\rm s} k,i}~=~{\rm e}/(T(1~-~i\tau/T))~=~\gamma_{{\rm s} k,0}/(1~-~i\tau/T)$. The denominator of (\ref{stconctra}) is proportional to $\Big({2^{\frac{\mathcal{B}}{TW\left(1-\frac{i\tau}{T}\right)}}-1} \Big) (1-i\frac{\tau}{T})$, which in turn monotonically decreasing with $i\tau$. Using the first derivative with respect to $i\tau$, the numerator of (\ref{stconctra}), $\overline{P_{{\rm s}k,i}}~=~\exp\big(\!-\!\frac{{2^{\frac{\mathcal{B}}{TW\left(1-\frac{i\tau}{T}\right)}}-1}}{\frac{\rm e}{T(1-i\frac{\tau}{T})} \sigma_{{\rm s} k}}\big)$, can be easily shown to be decreasing with $i\tau$ as in \cite{wimob,ElSh1312:Optimal}. Since the numerator of (\ref{stconctra}) is monotonically decreasing with $i\tau$ and the denominator is monotonically increasing with $i$, $ \overline{P_{{\rm s} k,i0}^{\left({\rm p}\right)}}$ is monotonically decreasing with $i\tau$. Therefore, the delay in the secondary access causes reduction in the probabilities of the secondary packets correct reception and the primary relayed packets correct reception at their destinations.

Now, we compute the ratio $\frac{\overline{P_{jk,1n}^{\left({v}\right)}}}{\overline{P_{jk,0}}}$. Using (\ref{conctra}), we have
%
%
 \begin{equation}
 \begin{split}
 \label{cfg2}
\hat{\delta}^{\left(v\right)}_{jk}\!=\! \frac{\overline{P_{jk,1}}}{\overline{P_{jk,0}}}&=\frac{\overline{P^{\left(v\right)}_{jk,1n}}}{\delta^{\left(v\right)}_{jk,1n}\overline{P_{jk,0}}}
 \end{split}
 \end{equation}
After some mathematical manipulations, the ratio $\frac{\overline{P_{jk,1n}^{\left({v}\right)}}}{\overline{P_{jk,0}}}$ is given by
  \begin{eqnarray}
\frac{\overline{P_{jk,1n}^{\left({v}\right)}}}{\overline{P_{jk,0}}}\!=\!{\delta^{\left(v\right)}_{jk,1n}}\hat{\delta}^{\left(v\right)}_{jk}
\end{eqnarray}
Note that throughout the paper, the superscript `$\left(v\right)$' can be eliminated from symbols since we only have two nodes: one PU and one SU. That is, ${\delta^{\left(v\right)}_{jk,in}}={\delta_{jk,in}}$, $\hat{\delta}^{\left(v\right)}_{jk}=\hat{\delta}_{jk}$ and $\overline{P_{jk,in}^{\left({v}\right)}}=\overline{P_{jk,in}}$.
\section*{Appendix B}

\begin{figure}
  \centering
  \includegraphics[width=1\columnwidth]{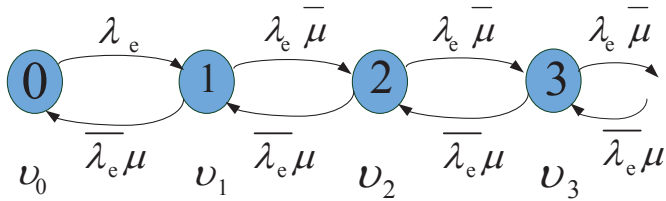}\\
  \caption{The Markov chain model for the secondary energy queue when its service rate is independent of the other queues and has a mean service rate $0\!\le\!\mu\!\le\! 1$.
  }\label{figenergy}
  \end{figure}

When the arrival and departure of the secondary energy queue become decoupled from all other queues in the network as in the approximated systems, we can construct and solve its Markov chain. The Markov chain is shown in Fig. \ref{figenergy}, where the mean arrival rate is $\lambda_{\rm e}$ and the mean service rate is $\mu$. Solving the state balance equations of the Markov chain modeling the secondary energy queue, it is straightforward to show that the probability that the energy queue has $1\le \vartheta\le\infty$ packets, $\nu_\vartheta$, is
\begin{equation}
\nu_\vartheta=\nu_{\circ}\frac{1}{\overline{\mu}}\Bigg(\frac{\lambda_{\rm e} \overline{\mu}}{\overline{\lambda_{\rm e} }\mu}\Bigg)^{\vartheta}=\nu_{\circ}\frac{\eta^{\vartheta}}{\overline{\mu}}, \ \vartheta=1,2,\dots,\infty
\end{equation}
where $\eta=\frac{\lambda_{\rm e} \overline{\mu}}{\overline{\lambda_{\rm e} }\mu}$.
Since the sum of all states' probabilities is the unity, $\sum_{\vartheta=0}^{\infty}\nu_\vartheta=1$. The probability of the secondary energy queue being empty is obtained via solving the following linear equation:
\begin{equation}
\nu_{\circ}+\sum_{\vartheta=1}^{\infty}\nu_\vartheta=\nu_{\circ}+\nu_{\circ}\sum_{\vartheta=1}^{{\infty}}\frac{1}{\overline{\mu}}\eta^{\vartheta}\!=\!1.
\end{equation}
After some mathematical manipulations, $\nu_{\circ}$ is given by
\begin{equation}
\nu_{\circ}\!=\!1-\frac{\lambda_{\rm e}}{\mu},
\end{equation}
with $\lambda_{\rm e}<\mu$. If $\lambda_{\rm e}\ge\mu$, the energy queue saturates, i.e., becomes always backlogged. Thus, $v_\circ=0$, which boosts the secondary rate. The probability of the primary energy queue being empty is $1-\lambda_{\rm e}/{\mu}$.

If $\mu=1$, $\eta=0$ and the probability that the energy queue having more than one packet is zero. The states probabilities in such case are given by
\begin{equation}
\nu_0=1-\lambda_{\rm e}, \ \nu_1=\lambda_{\rm e}, \ \nu_\vartheta=0, \ \vartheta=2,\dots,\infty
\end{equation}

\bibliographystyle{IEEEtran}
\bibliography{IEEEabrv,term_bib}

\begin{thebibliography}{10}
\providecommand{\url}[1]{#1}
\csname url@samestyle\endcsname
\providecommand{\newblock}{\relax}
\providecommand{\bibinfo}[2]{#2}
\providecommand{\BIBentrySTDinterwordspacing}{\spaceskip=0pt\relax}
\providecommand{\BIBentryALTinterwordstretchfactor}{4}
\providecommand{\BIBentryALTinterwordspacing}{\spaceskip=\fontdimen2\font plus
\BIBentryALTinterwordstretchfactor\fontdimen3\font minus
  \fontdimen4\font\relax}
\providecommand{\BIBforeignlanguage}[2]{{%
\expandafter\ifx\csname l@#1\endcsname\relax
\typeout{** WARNING: IEEEtran.bst: No hyphenation pattern has been}%
\typeout{** loaded for the language `#1'. Using the pattern for}%
\typeout{** the default language instead.}%
\else
\language=\csname l@#1\endcsname
\fi
#2}}
\providecommand{\BIBdecl}{\relax}
\BIBdecl

\bibitem{myprotocol}
A.~{El Shafie} and T.~Khattab, ``Maximum throughput of a cooperative energy
  harvesting cognitive radio user,'' { Accepted} in {\it Personal, Indoor and
  Mobile Radio Communications (PIMRC)}, 2014.

\bibitem{survey}
S.~Sudevalayam and P.~Kulkarni, ``Energy harvesting sensor nodes: Survey and
  implications,'' \emph{IEEE Communications Surveys and Tutorials}, vol.~13,
  no.~3, pp. 443--461, 2011.

\bibitem{lei2009generic}
J.~Lei, R.~Yates, and L.~Greenstein, ``A generic model for optimizing
  single-hop transmission policy of replenishable sensors,'' \emph{IEEE Trans.
  Wireless Commun.}, vol.~8, no.~2, pp. 547--551, 2009.

\bibitem{sharma2010optimal}
V.~Sharma, U.~Mukherji, V.~Joseph, and S.~Gupta, ``Optimal energy management
  policies for energy harvesting sensor nodes,'' \emph{IEEE Trans. Wireless
  Commun.}, vol.~9, no.~4, pp. 1326--1336, 2010.

\bibitem{gatzianas2010control}
M.~Gatzianas, L.~Georgiadis, and L.~Tassiulas, ``Control of wireless networks
  with rechargeable batteries [transactions papers],'' \emph{IEEE Trans.
  Wireless Commun.}, vol.~9, no.~2, pp. 581--593, 2010.

\bibitem{pappas2012optimal}
N.~Pappas, J.~Jeon, A.~Ephremides, and A.~Traganitis, ``Optimal utilization of
  a cognitive shared channel with a rechargeable primary source node,'' in
  \emph{JCN}, vol.~14, no.~2, Apr. 2012, pp. 162--168.

\bibitem{wimob}
A.~El~Shafie and A.~Sultan, ``Optimal random access and random spectrum sensing
  for an energy harvesting cognitive radio,'' in \emph{Proc. IEEE 8th WiMob},
  2012, pp. 403--410.

\bibitem{ourletter}
A.~{El Shafie} and A.~Sultan, ``Optimal random access for a cognitive radio
  terminal with energy harvesting capability,'' \emph{IEEE Communications
  Letters}, vol.~17, no.~6, pp. 1128--1131, 2013.

\bibitem{ElSh1312:Optimal}
------, ``Optimal selection of spectrum sensing duration for an energy
  harvesting cognitive radio,'' in \emph{IEEE GLOBECOM}, Dec 2013, pp.
  1020--1025.

\bibitem{krikidis2012stability}
I.~Krikidis, T.~Charalambous, and J.~Thompson, ``Stability analysis and power
  optimization for energy harvesting cooperative networks,'' \emph{IEEE Signal
  Process. Lett.}, vol.~19, no.~1, pp. 20--23, 2012.

\bibitem{wcmpaper}
A.~{El Shafie}, T.~Khattab, A.~El-Keyi, and M.~Nafie, ``On the coexistence of a
  primary user with an energy harvesting secondary user: A case of cognitive
  cooperation,'' 2013, {To} appear in {\it Wireless Communications and Mobile
  Computing}.

\bibitem{sadek}
A.~Sadek, K.~Liu, and A.~Ephremides, ``Cognitive multiple access via
  cooperation: protocol design and performance analysis,'' \emph{IEEE Trans.
  Inf. Theory}, vol.~53, no.~10, pp. 3677--3696, Oct. 2007.

\bibitem{simeone}
O.~Simeone, Y.~Bar-Ness, and U.~Spagnolini, ``Stable throughput of cognitive
  radios with and without relaying capability,'' \emph{IEEE Trans. Commun.},
  vol.~55, no.~12, pp. 2351--2360, Dec. 2007.

\bibitem{krikidis2009protocol}
I.~Krikidis, J.~Laneman, J.~Thompson, and S.~McLaughlin, ``Protocol design and
  throughput analysis for multi-user cognitive cooperative systems,''
  \emph{IEEE Trans. Wireless Commun.}, vol.~8, no.~9, pp. 4740--4751, Sept.
  2009.

\bibitem{krikidis2010stability}
I.~Krikidis, N.~Devroye, and J.~Thompson, ``Stability analysis for cognitive
  radio with multi-access primary transmission,'' \emph{IEEE Trans. Wireless
  Commun.}, vol.~9, no.~1, pp. 72--77, Jan. 2010.

\bibitem{close}
J.~Gambini, O.~Simeone, and U.~Spagnolini, ``Cognitive relaying and
  opportunistic spectrum sensing in unlicensed multiple access channels,'' in
  \emph{IEEE 10th ISSSTA}, Aug. 2008, pp. 371--375.

\bibitem{rao1988stability}
R.~Rao and A.~Ephremides, ``On the stability of interacting queues in a
  multiple-access system,'' \emph{IEEE Trans. Inf. Theory}, vol.~34, no.~5, pp.
  918--930, Sept. 1988.

\bibitem{luo1999stability}
W.~Luo and A.~Ephremides, ``Stability of {N} interacting queues in
  random-access systems,'' \emph{IEEE Trans. Inf. Theory}, vol.~45, no.~5, pp.
  1579--1587, July 1999.

\end{thebibliography}
\balance
\end{document}